\documentclass{emulateapj}
\usepackage{apjfonts}

\def\Mo{\rm{M}_\odot}

\def\note #1]{\noindent{\bf #1]}}

\def\sgnk2{{\rm sgn\left(K^2\right)}}
\def\rmd{{\rm d}}
\def\rmdd{{\rm d}^2}
\def\mF{\mathcal{F}}

\newcommand{\mitbf}[1] {\hbox{\mathversion{bold}$#1$}}

\shorttitle{ASTEROSEISMIC SIGNATURES OF CONVECTIVE CORES}
\shortauthors{CUNHA \& METCALFE}

\begin{document}
 
\title{Asteroseismic Signatures of Small Convective Cores}

\author{M.~S. Cunha\altaffilmark{1}}
\affil{Centro de Astrof{\'\i}sica da Universidade do Porto, Rua das 
Estrelas, 4150-Porto, Portugal}
\altaffiltext{1}{Visiting Scientist, High Altitude Observatory, NCAR, PO 
Box 3000, Boulder, CO 80307 USA}

\and

\author{T.~S. Metcalfe}
\affil{High Altitude Observatory and Scientific Computing Division, NCAR, 
PO Box 3000, Boulder, CO 80307 USA}

\email{mcunha@astro.up.pt, travis@hao.ucar.edu}
\slugcomment{Astrophysical Journal, accepted for publication}

\begin{abstract}

We present an analytical study of the effect of small convective cores on 
the oscillations of solar-like pulsators. Based on an asymptotic analysis 
of the wave equation near the center of the star, we derive an expression 
for the perturbations to the frequencies of radial modes generated by a 
convective core and discuss how these perturbations depend on the 
properties of the core. Moreover, we propose a diagnostic tool to isolate 
the predicted signature of the core, constructed from a particular 
combination of the oscillation frequencies, and we validate this tool with 
simulated data. We also show that the proposed tool can be applied to the 
pulsation data soon expected from satellite missions such as CoRoT and 
Kepler to constrain the amplitude of the discontinuity in the sound speed 
at the edge of the convective core, the ratio between the sound speed and 
the radius at this same location, and the stellar age.

\end{abstract}


%%%%%%%%%%%%%%%%%%%%%%%%%%%%%%%%%%%%%%%%%%%%%%%%%%%%%%%%%%%%%%%%%%%%%%%%%%%

\section{Introduction}

Space-based asteroseismology began in 1999, when the 5\,cm star tracker 
mounted on the Wide-field InfraRed Explorer (WIRE) satellite was used to 
detect oscillations in the K giant $\alpha$~UMa \citep{buz00}, and this 
instrument has continued to produce a steady stream of asteroseismic data 
\citep[e.g., see][]{bru07}. The 15\,cm Micro-variability \& Oscillations 
of Stars (MOST) satellite was launched in 2003, and was the first 
satellite in orbit that was actually designed for asteroseismology 
\citep{wal03}. Although MOST's failure to detect solar-like oscillations 
in Procyon \citep{mat04} was later traced to a larger than expected source 
of non-white instrumental noise \citep{bed05}, its ability to conduct 
nearly uninterrupted time-series photometry for durations of up to two 
months has produced exquisite data for pulsating stars across the H-R 
diagram \citep[e.g., see][]{ran05,aer06,sai06,row07,cam07}.

The next era of space-based asteroseismology began with the successful 
launch of the 27\,cm Convection, Rotation and planetary Transits (CoRoT) 
satellite in December 2006 \citep{bag06}, and will continue with the 
anticipated launch of the 95\,cm Kepler satellite in 2008 \citep{jcd07}. 
These missions promise advances in the study of solar-like oscillations 
comparable to what WIRE and MOST have achieved for the larger amplitude 
classical pulsators. With the frequency precision expected to approach 
$\sim$0.1~$\mu$Hz from this next generation of satellites, we can begin to 
consider the possible detection of extremely subtle asteroseismic signals.

There is a long history of constructing combinations of the observed 
oscillation frequencies to extract useful asteroseismic information. The 
simplest such combination is the average frequency spacing between modes 
of consecutive radial order but the same spherical degree---{\it the large 
separation}---which provides an estimate of the mean stellar density 
\citep{cox80}. Similarly, the average spacing between modes of consecutive 
radial order but with spherical degrees that differ by 2---{\it the small 
separation}---is sensitive to chemical gradients in the deep interior, 
tracing the stellar evolutionary state \citep{dap88}.

The idea that sharp variations in the internal structure of distant stars 
generate signatures in the frequencies of low degree modes that can be 
recognized and interpreted with the appropriate seismic tools has been 
explored in a number of papers. In this context, theoretical derivations 
were carried out of the signal expected from sharp variations taking place 
at the boundary of convective envelopes \citep{monteiro00} and in the 
region of helium ionization \citep{monteiro98,houdek07}. Moreover, the 
expected signature of convective cores on the small frequency separations 
has also been considered by \citet{roxburgh01}. However, all of these 
works assumed that the sharp variation in the structure under 
investigation was located far from the turning points of the oscillation 
modes, i.e., well within their propagation cavities.  Consequently, the 
analyses carried out in these works do not directly apply to studies of 
small convective cores, such as those present in main-sequence solar-like 
pulsators.

Attempts to identify and interpret in a systematic manner the signatures 
of convective cores in solar-like pulsators directly from simulated data 
have recently been carried out by \citet{maz06}. Based on an analysis of 
the results obtained from simulated data, the authors suggested a 
diagnostic tool to estimate the masses of the convective cores and the 
stellar ages. While recognizing the importance of studying simulated data, 
the chances of constructing the optimal diagnostic tool to detect and 
characterize small convective cores may be greater if we know what signal 
to expect. Moreover, a theoretical understanding of the expected signal 
can provide us with a direct link between the structure under 
investigation and the observations---moving us toward our goal of finding 
the best way to infer information about that structure. With this in mind, 
in this paper we derive the expected signal of a small convective core on 
the oscillation frequencies of solar-like pulsators. Unlike 
\citet{monteiro00}, \citet{roxburgh01}, and \citet{houdek07}, we do not 
assume that the region of sharp structural variation, which in the present 
case is the edge of the convective core, is placed well within the 
propagation region of the modes. In fact, the analysis is focused on the 
opposite case, in which the edge of the convective core is sufficiently 
close to the center to affect the radial modes more significantly than the 
modes with degree $\ell\ge 1$.

We derive the theoretical expression for the signature of a small 
convective core in \S\ref{theory}. In \S\ref{tool} we construct a 
diagnostic tool, based on a particular combination of the frequencies, to 
isolate this signature. In \S\ref{DATA} we validate the theoretical 
predictions and the diagnostic tool using simulated data from a series of 
1.3~$\Mo$ models. Moreover, we discuss the successes and limitations of 
the proposed tool, as well as the information that it can reveal. Finally, 
in \S\ref{DISC} we outline the potential for future work in this area, and 
we discuss possible observational tests.

%%%%%%%%%%%%%%%%%%%%%%%%%%%%%%%%%%%%%%%%%%%%%%%%%%%%%%%%%%%%%%%%%%%%%%%%%%%

\section{Signature of a convective core}

\subsection{\label{theory}Theoretical expectation}

Let ${\mitbf \xi}=(\xi,0,0)r\, e^{-i\omega t}$ be the displacement, 
defined with respect to a spherical coordinate system $(r,\theta,\phi)$, 
for radial, adiabatic, acoustic oscillations in a spherically symmetric 
star, with $\omega$ the oscillation frequency and $t$ the time. Through 
the appropriate mathematical manipulation of the equations of motion and 
mass conservation for radial pulsations, under the adiabatic 
approximation, the dimensionless displacement $\xi$ is found to obey the 
second order differential equation \citep[e.g.][]{gough93},
\begin{equation}
\frac{{\rm d}}{{\rm d}r}\left(\gamma p r^4\frac{{\rm d}\xi}{{\rm 
d}r}\right)+ \left\{r^3\frac{\rm d}{{\rm d}r}\left[\left(3\gamma 
-4\right)p\right]+r^4\rho\omega^2\right\}\xi =0,
\label{xi}
\end{equation}
where $\gamma$ is the first adiabatic exponent, $p$ is the pressure and 
$\rho$ is the density. By defining new dependent and independent 
variables, $\Xi=r_{\rm ref}^{1/2}r^{3/2}\rho^{1/2}c\,\xi$ and 
$z=\ln(r/r_{\rm ref})$ respectively, where $c=\sqrt{\gamma p/\rho}$ is the 
sound speed and $r_{\rm ref}$ is a fiducial value of $r$, and substituting 
into Eq.~(\ref{xi}), the latter reduces to the standard form,
\begin{equation}
\frac{{\rm d}^2 \Xi}{{\rm d}z^2}+K^2\Xi=0,
\label{waveeq}
\end{equation}
where
\begin{equation} 
K^2=\frac{\omega^2-\omega_{\rm c}^2}{c^2}\,r^2-\frac{1}{4},
\nonumber
\end{equation}
and $\omega_{\rm c}$ is a critical acoustic frequency \citep[e.g., 
see][Eq.~4.8.7]{gough93}. Since our aim is to study the effect of a small 
convective core on the eigenfrequencies, we are particularly interested in 
the form of $\omega_c$ close to the center of the star. From the general 
expression for $\omega_{\rm c}$, we find that when $r\rightarrow 0$, 
$\omega_{\rm c}^2\approx 2c^2/r^2$. Using this expression for $\omega_{\rm 
c}^2$, we then find the critical frequency $\omega_{\rm c}^*$ at which 
$K^2=0$ to be,
\begin{equation}
\omega_{\rm c}^*\approx\frac{3\,c}{2\,r}.
\label{wc}
\end{equation}

\begin{figure}
\plotone{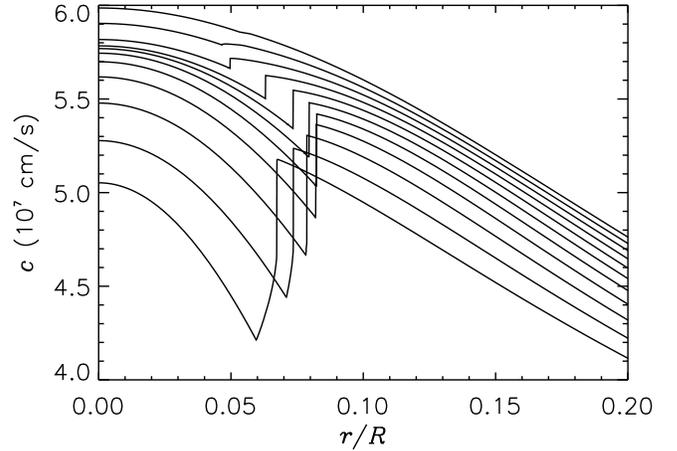}
\caption{Sound speed in the inner layers of main-sequence models with 
$M=1.3\,\Mo$, and ages ranging from 0.25~Gyr (top curve) to 5.25~Gyr 
(bottom curve) in intervals of 0.5~Gyr. The discontinuity in $c$ marks the 
edge of the convective core, which expands during the evolution of the 
star up to an age of 3.25~Gyr and then begins to contract.}
\label{sspeed}
\end{figure}

A model with a small convective core exhibits a sharp structural variation 
close to the inner turning point of Eq.~(\ref{waveeq}). This sharp 
variation can be seen in the sound speed profiles of main-sequence models 
with mass $M=1.3\,\, \Mo$, shown in Figure~\ref{sspeed} for a range of 
stellar ages. Our aim is to determine how the eigenfrequencies of an 
otherwise similar model with a smooth structure around the edge of the 
convective core---hereafter the {\it unperturbed model}---are modified by 
the presence of this sharp variation.

% FIGURE 1 %%%%%%%%%%%%%%%%%%%%%%%%%%%%%%%%%%%%%%%%%%%%%%%%%%%%%%%%%%%%%%
%%%%%%%%%%%%%%%%%%%%%%%%%%%%%%%%%%%%%%%%%%%%%%%%%%%%%%%%%%%%%%%%%%%%%%%%%

\begin{figure} 
\plotone{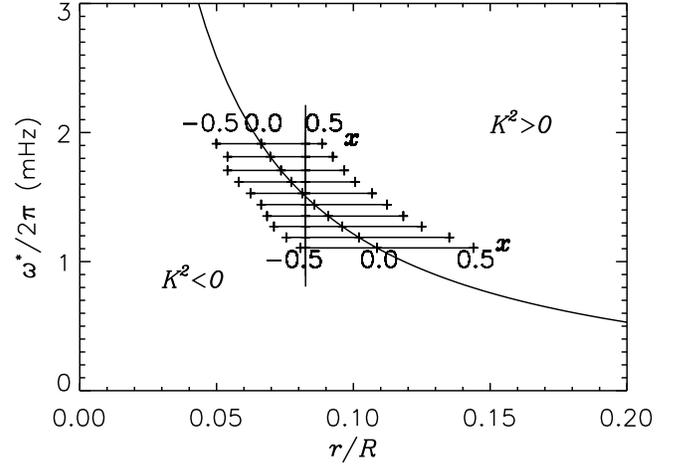} 
\caption{The full curve shows the
critical frequency obtained by imposing $K^2=0$ for the inner layers
of an unperturbed main-sequence model with $M=1.3\, \Mo$ at the age
when its convective core reaches maximum size (3.25~Gyr). For each
frequency, this curve gives the maximum propagation depth of a radial
mode when the eigenfunctions are the solutions to
Eq.~(\ref{waveeq}). The vertical line shows the radial position of the
edge of the convective core. The horizontal lines are axes for the
independent variable $x$ in the range $[-0.5,0.5]$. These are shown for 
eigenmode frequencies of the corresponding perturbed model, with radial 
orders ranging from $n=11$ (bottom line) to $n=20$ (top line).}
\label{pd} 
\end{figure}

To achieve this we must find an expression for $\Xi$ that is valid close 
to the inner turning point of Eq.~(\ref{waveeq}). Such an expression for 
$\Xi$ is derived in Appendix \ref{APA}. In short, we define new dependent 
and independent variables and substitute them into Eq.~(\ref{waveeq}) to 
obtain an equation that resembles the Airy equation with an additional 
term (cf.~Eq.~\ref{airy} in Appendix~\ref{APA}). Applying Olver's 
comparison method \citep{olver74} we then find, close to the inner turning 
point,
\begin{eqnarray}
\Xi \approx a \,|x|^{1/4}|K|^{-1/2}\left[A_{\rm 
i}\left(0\right)- {A^\prime}_{\rm i}\left(0\right)x\right],
\label{xiapprox}
\end{eqnarray}
where $a$ is a constant related to the normalization of 
eigenfunctions, $A_{\rm i}(0)$ and ${A^\prime}_{\rm i}(0)$ are the values 
taken at $x=0$ of the first solution to the Airy equation and its first 
derivative respectively, and $x$ is the new independent variable, which is 
related to $z$ through Eq.~(\ref{defx}) in Appendix~\ref{APA}. In deriving 
Eq.~(\ref{xiapprox}) we have neglected terms of $\mathcal{O}(x^3)$. So, 
this equation is approximately correct when $x$ is sufficiently close to 
0. Note that unlike $r$, the independent variable $x$ depends on the value 
of the eigenfrequency. For a given frequency it goes through 0 at the 
inner turning point, and is negative when $K^2 < 0$ and positive when $K^2 
> 0$. The relation between $x$ and $r$ for different values of the 
eigenfrequency $\omega$, as well as the regions where $K^2 <0$ and 
$K^2>0$, are shown in Figure~\ref{pd} for an unperturbed stellar model 
with $M=1.3\,\,\Mo$. The unperturbed models were obtained by fitting the 
 sound speed profiles of each stellar model with smooth 
functions around the edge of the convective core. The density profile was 
then derived by assuming hydrostatic equilibrium. 

To study the effect of a small convective core on the oscillation 
frequencies, we start from an integral equation for the frequency, which 
can be derived from Eq.~(\ref{xi}) after multiplying by $\xi$ and 
integrating by parts between $r=0$ and $r=R$, where $R$ is the radius at 
the stellar surface. The resulting well known equation is,
\begin{equation}
\omega^2=\frac{I_2-{\mathcal B}}{I_1},
\label{vp}
\end{equation}
where
\begin{eqnarray}
I_2 & = &\int_0^R{\left[\gamma pr^4\left(\frac{d\xi}{dr}\right)^2-
r^3\frac{d}{dr} \left[\left(3\gamma-4\right)p\right] \xi^2\right]dr}, 
\nonumber\\
I_1 & = & \int_0^R{\left[\rho r^4\xi^2\right]dr}, 
\nonumber
\end{eqnarray}
and
\begin{eqnarray}
{\mathcal B} & = & {\mathcal B}(R)-{\mathcal B}(0) = 
\left[\gamma pr^4\xi\frac{d\xi}{dr}\right]_{r=0}^{r=R}. 
\nonumber
\end{eqnarray}
When the oscillations are below the acoustic cutoff frequency, the surface 
term ${\mathcal B}$ can safely be neglected. In fact, it follows from the 
regularity condition at $r=0$ that ${\mathcal B}(0)=0$ and, given the 
small value of $p(R)$, if the oscillations are evanescent in the outer 
layers ${\mathcal B}(R)$ will also be small.

We now consider the above mentioned unperturbed smooth model and an 
otherwise similar model with a sharp structural variation at the edge of a 
small convective core---hereafter the {\it perturbed model}. If ${\mathcal 
B}$ is neglected in Eq.~(\ref{vp}), the difference between the oscillation 
frequencies in the perturbed and unperturbed models in a linear 
approximation is given by
\begin{equation}
\delta\omega=\frac{\delta I_2-\omega^2\delta I_1}{2\omega I_1},
\label{vpp}
\end{equation}
where all unperturbed quantities refer to the smooth model. Equation 
(\ref{vp}) with ${\mathcal B}$ neglected constitutes a variational 
principle for the frequency $\omega$ 
\citep[e.g.][]{chandrasekhar64,gough93}. Thus, to first order the 
perturbation to the eigenfunctions does not contribute to the perturbation 
to the eigenfrequencies and we can write,
\begin{eqnarray}
\delta I_1 = \int_0^R \delta\rho r^4\xi^2{\rm dr},
\label{di1}
\end{eqnarray} 
and
\begin{eqnarray}
\delta I_2 & = & \int_0^R\left\{r^4\delta\left(\gamma p\right)
\left(\frac{{\rm d}\xi}{{\rm d}r}\right)^2\right. 
\nonumber\\ & & 
\left.-r^3\frac{\rm d}{{\rm d}r}\left[3\delta\left(\gamma p\right)
-4\delta p\right]\xi^2\right\}{\rm d}r,
\label{di2}
\end{eqnarray}
where $\delta\rho$, $\delta p$ and $\delta\gamma$ denote perturbations at 
fixed radius.
To obtain an expression for $\delta\omega$ as a function of $\omega$, we 
manipulate Eqs.~(\ref{vpp}-\ref{di2}) in a way very similar to that 
presented in \citet{monteiro94}. In particular, Eqs.~(\ref{di1}) and 
(\ref{di2}) are written in terms of the perturbations $\delta (\gamma p)$, 
$\delta c^2$ and $\delta g$, with $g=-\frac{1}{\rho} \frac{{\rm d}p}{{\rm 
d}r}$ and then integrated by parts (see Appendix~\ref{APB} for details). 
As discussed in Appendix~\ref{APB}, the perturbation to the 
eigenfrequencies is determined predominantly by the terms associated with 
$\delta c^2$.  Describing this perturbation with a modified step function, 
we find (cf.~Eq.~\ref{final})
\begin{eqnarray}
\delta\omega \approx -\frac{A_\delta}{2\omega I_1}\left[\frac{|K|r^3\rho 
c^2\xi^2}{|x|^{1/2}}\right]_{x=0}\left[\frac{r\left(4g+\omega^2r\right)}
{c^2|K|^2}\right]_{x=x_{\rm d}} 
\nonumber\\ 
\times \left[|x|x-\frac{A^\prime_{\rm i}(0)}{A_{\rm i}(0)}|x|x^2+\frac{1}{3}
\frac{A^\prime_{\rm i}(0)^2}{A_{\rm i}(0)^2}|x|x^3\right]_{x=x_{\rm d}},
\label{dw2}
\end{eqnarray}
where $A_\delta$ is a positive constant related to the sound speed 
increase at the edge of the convective core and $x_{\rm d}$ is the value 
taken by the independent variable $x$ at the location of the edge of the 
convective core. Both $x$ and $x_{\rm d}$ are functions of frequency. 
Through this dependence, Eq.~(\ref{dw2}) expresses how the frequency 
perturbation is modulated with frequency. According to this equation the 
frequency perturbation goes through zero when $x_{\rm d}=0$, i.e.~when the 
frequency is such that the turning point and the edge of the convective 
core are located at the same depth.

% FIGURE 2 %%%%%%%%%%%%%%%%%%%%%%%%%%%%%%%%%%%%%%%%%%%%%%%%%%%%%%%%%%%%%%
%%%%%%%%%%%%%%%%%%%%%%%%%%%%%%%%%%%%%%%%%%%%%%%%%%%%%%%%%%%%%%%%%%%%%%%%%

If the signal anticipated by Eq.~(\ref{dw2}) could be detected in the 
oscillation frequencies of real stars, then information about the size and 
position of the sound speed discontinuity at the edge of the convective 
core could be inferred from the amplitude $A_\delta$ and the frequency at 
which $\delta\omega=0$, respectively. With this in mind, in the next 
section we present a diagnostic tool that will allow us to isolate the 
signal derived above.

\subsection{\label{tool}Diagnostic tool}

When deriving Eq.~(\ref{dw2}) we assumed that our model differs from a 
smooth model only at the edge of the convective core. However, sharp 
variations in the internal structure of a star are known to occur also at 
other locations, such as the base of the outer convective region and the 
region of helium ionization. These, in turn, will generate additional 
perturbations to the eigenfrequencies, as compared with those of a smooth 
model \citep[e.g.,][]{monteiro94,basu97,monteiro05,houdek07}. Thus, to 
isolate the signal produced by the edge of a convective core, we must 
combine the oscillation frequencies of real stars in such a way as to 
cancel out, as much as possible, the signal coming from all other sources. 
With this in mind, we write the frequency 
$\nu_{n,\ell}=\omega_{n,\ell}/2\pi$ of a mode with radial order $n$ and 
degree $\ell$ in a given star as the sum of three components
\begin{equation}
\nu_{n,\ell} = \nu_{n,\ell}^s+\delta\nu_{n,\ell}^p+\delta\nu_{n,\ell}^c,
\label{mu}
\end{equation}
where $\nu_{n,\ell}^s$ is the frequency of the mode in the smooth model, 
$\delta\nu_{n,\ell}^p$ is the perturbation to that frequency arising from 
all sharp variations in the structure of the star taking place well within 
the propagation region of the same mode (which we shall consider to be of 
low degree, with $\ell\le 3$), and $\delta\nu_{n,\ell}^c$ is the frequency 
perturbation produced by the edge of the small convective core.

While sharp variations in the envelope of a star will affect the 
frequencies of all modes of low degree, the discontinuity in the sound 
speed at the edge of a small convective core for a star slightly more 
massive than the Sun might be expected to affect primarily the frequencies 
of radial modes. This is not to say that the frequencies of modes of 
degree $\ell=1$, which also propagate to very deep layers, will not be 
affected at all by the convective core. In fact, to correctly derive the 
expected signal for a dipolar mode, the Cowling approximation should be 
avoided. The derivation should then be started from the formalism recently 
presented by \citet{takata06}, which shows that the inner turning point of 
dipolar modes is modified when the equations are treated without 
neglecting the perturbation to the gravitational potential. However, such 
an analysis is beyond the scope of the present paper. Hence, based on the 
idea that radial modes are most sensitive to the deepest layers of a star, 
we will proceed by assuming $\delta\nu_{n,\ell}^c\approx 0$ for modes with 
$\ell\ge 1$. This assumption will be tested in the following section. 
Moreover, given that we are interested in modes with degree $\ell\leq 3$, 
we neglect the weak dependence of $\delta\nu_{n,\ell}^p$ on $\ell$ 
\citep[e.g.,][]{monteiro94,houdek07} when subtracting the frequencies of 
nearly degenerate modes. With these assumptions, the {\it scaled small 
separations} 
$D_{\ell,\ell+2}\equiv\left(\nu_{n,\ell}-\nu_{n-1,\ell+2}\right)/ 
\left(4\ell+6\right)$ for pairs of modes with degree $\ell\leq 3$ divided 
by the large separations 
$\Delta\nu_{n,\ell}\equiv\nu_{n+1,\ell}-\nu_{n,\ell}$ take the form
\begin{eqnarray}
\frac{D_{02}}{\Delta\nu_{n-1,1}}& \approx &
\frac{\nu_{n,0}^s-\nu_{n-1,2}^s}{6\,{\Delta\nu_{n-1,1}}}+
\frac{\delta\nu_{n,0}^c}{6\,{\Delta\nu_{n-1,1}}},
\label{ss02}\\
\frac{D_{13}}{\Delta\nu_{n,0}}& \approx &
\frac{\nu_{n,1}^s-\nu_{n-1,3}^s}{10\,{\Delta\nu_{n,0}}}
\label{ss13}.
\end{eqnarray}
In the high-frequency asymptotic regime, the quantity 
\begin{equation}
D_\delta\equiv\frac{\pi}{4\ell+6}\frac{\nu_{n,\ell}^s-\nu_{n-1,\ell+2}^s}
{{\Delta\nu_{n,\ell}}}
\label{bigd}
\end{equation}
in the unperturbed model is essentially independent of the pair of modes 
considered \citep[$\ell=0,2$ or $\ell=1,3$; 
see][]{roxburgh00b,roxburgh00a,roxburgh03}. Thus, we proceed by 
subtracting Eq.~(\ref{ss13}) from Eq.~(\ref{ss02}) to find the relation
\begin{equation}
\frac{D_{02}}{\Delta\nu_{n-1,1}}-\frac{D_{13}}{\Delta\nu_{n,0}}\approx
\frac{\delta\nu_{n,0}^c}{6\,{\Delta\nu_{n-1,1}}}.
\label{signal}
\end{equation}
Note that for the large separations appearing in the denominators, we used 
pairs of modes that encompass the frequency range of the modes used in the 
scaled small separations, which appear in the numerators. When defined in 
this way, the frequency differences presented in Eq.~(\ref{signal}) are a 
smoother function of frequency than when $\Delta\nu_{n,0}$ and 
$\Delta\nu_{n,1}$ are used in the denominators of the first and second 
term, respectively. In practice, this means that when subtracting 
Eq.~(\ref{ss13}) from Eq.~(\ref{ss02}) we also had to assume that 
$\frac{\Delta\nu_{n,0}}{\Delta\nu_{n-1,1}}\approx 
\frac{\Delta\nu_{n,1}}{\Delta\nu_{n,0}}$.

The frequency perturbation $\delta\nu_{n,0}^c$ is expected to be modulated 
with frequency according to Eq.~(\ref{dw2}). Thus, if the assumptions 
leading to Eq.~(\ref{signal}) hold, we would expect to see a corresponding 
modulation when the oscillation frequencies observed in a given solar-like 
pulsator are combined in the manner prescribed. In the next section we 
test this prediction on simulated data and we discuss what might be 
inferred about solar-like stars from this diagnostic tool, once 
observations with the necessary precision become available.

%%%%%%%%%%%%%%%%%%%%%%%%%%%%%%%%%%%%%%%%%%%%%%%%%%%%%%%%%%%%%%%%%%%%%%%%%%%

\section{\label{DATA}Tests with simulated data}

\subsection{\label{models}Models}

We used the Aarhus stellar evolution code \citep[ASTEC;][]{jcd82} coupled 
with the adiabatic pulsation code (ADIPLS) to calculate the theoretical 
oscillation frequencies for 1.3~$\Mo$ models with ages from 0.25 to 5.25 
Gyr, sampling the full range of convective core sizes along the evolution 
track. The ASTEC code used the equation of state (EOS) of \cite{eff73} 
without Coulomb corrections, and opacities from the OPAL tables 
\citep{ir96}, supplemented by Kurucz opacities at low temperatures. The 
nuclear reaction rates came from \cite{bp92}, convection was described by 
the mixing-length theory of \cite{bv58}, and we did not include the 
effects of diffusion.

We set the initial metallicity to $Z_0=0.02$, and the initial hydrogen 
mass fraction to $X_0=0.74$. We fixed the mixing-length parameter at 
$\alpha=1.9$ and included core overshoot using $\alpha_{\rm ov}=0.25$ with 
complete mixing in the overshoot region and considering the associated 
changes to the adiabatic temperature gradient. For each selected model, we 
calculated the radial and non-radial p-mode frequencies for oscillations 
with spherical degree $\ell$=0-3 and radial order $n$=1$\sim$30. We used 
these frequencies to construct the combinations shown in 
Eq.~(\ref{signal}), and compared them to the theoretical predictions.

% FIGURE 3 %%%%%%%%%%%%%%%%%%%%%%%%%%%%%%%%%%%%%%%%%%%%%%%%%%%%%%%%%%%%%%
\begin{figure}
\plotone{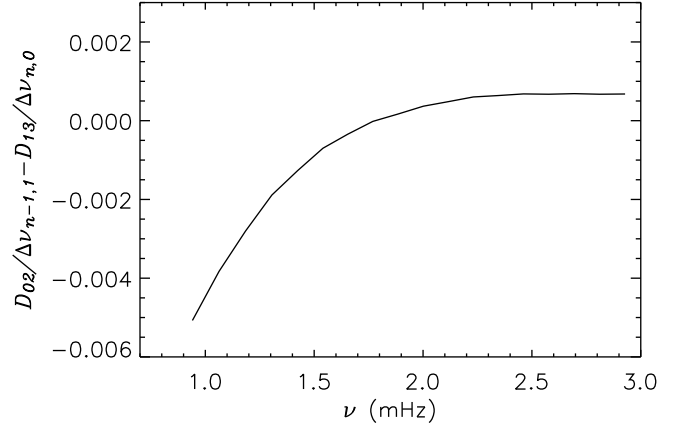}
\caption{Frequency differences, as defined by the left hand side of 
Eq.~(\ref{signal}), for a model with $M=1.3$~M$_\odot$ at an age of 0.25 
Gyr when its convective core is very small and there is no discontinuity 
in the sound speed (see the top curve in Figure~\ref{sspeed}).}
\label{difnocc}
\end{figure}
%%%%%%%%%%%%%%%%%%%%%%%%%%%%%%%%%%%%%%%%%%%%%%%%%%%%%%%%%%%%%%%%%%%%%%%%%

\subsection{\label{tests}Testing the diagnostic tool}

A number of approximations were made when deriving the diagnostic tool 
proposed in Eq.~(\ref{signal}). Thus, it is essential that we evaluate the 
ability of this seismic tool when it comes to isolating the signal from 
the convective core.

% FIGURE 4 %%%%%%%%%%%%%%%%%%%%%%%%%%%%%%%%%%%%%%%%%%%%%%%%%%%%%%%%%%%%%%
\begin{figure*}
\plottwo{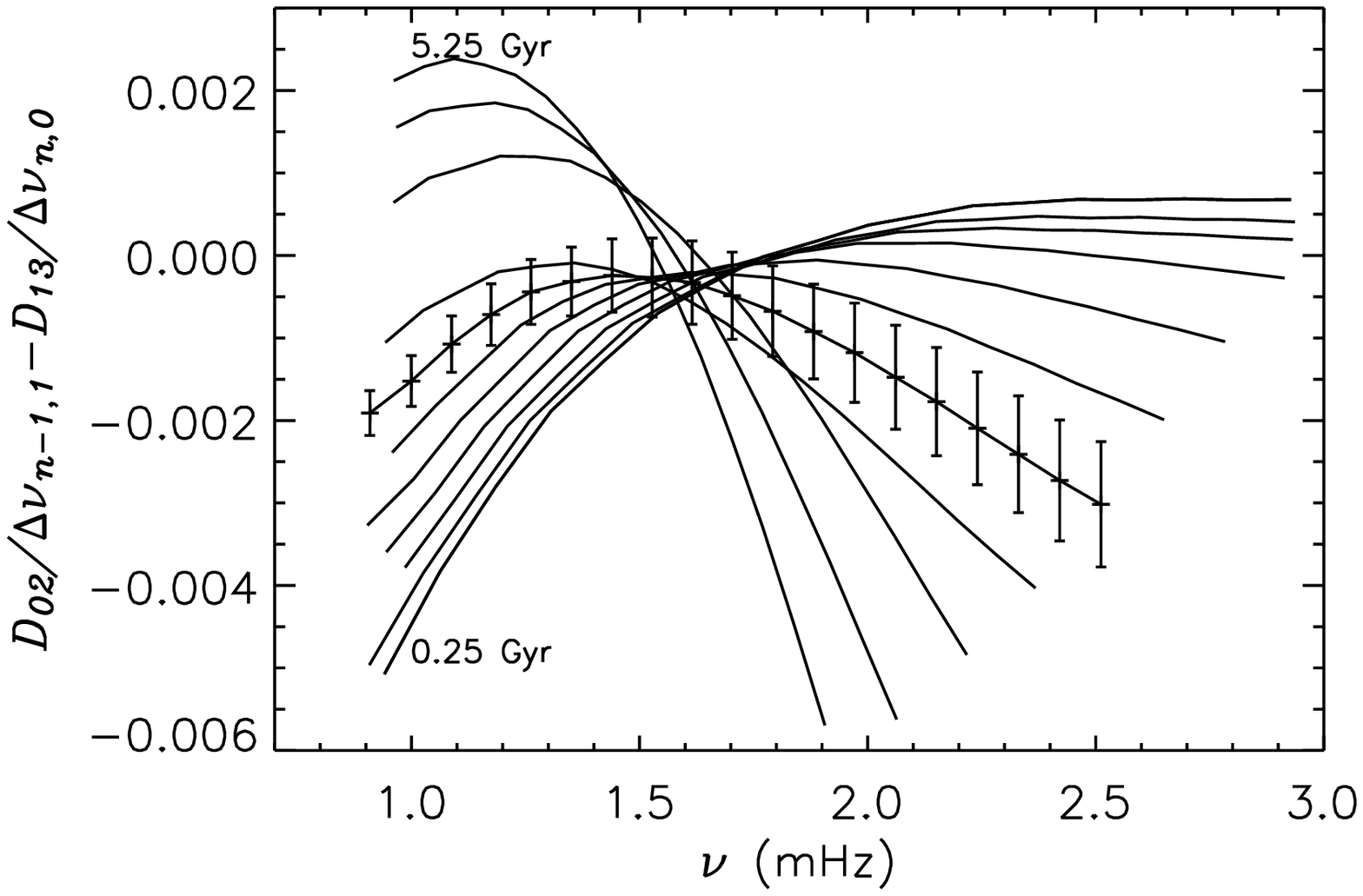}{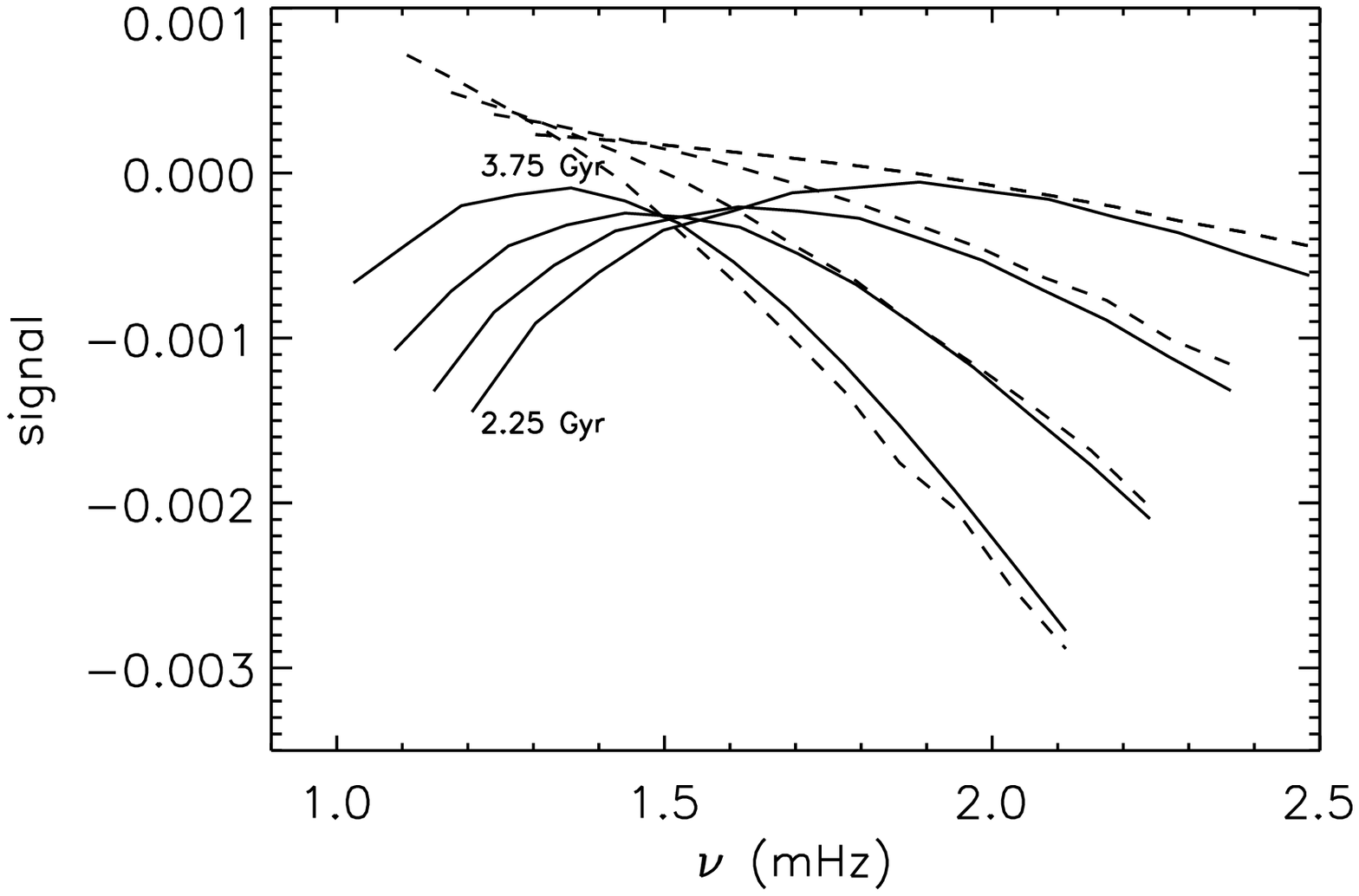}
\caption{Left panel: the same as Figure~\ref{difnocc}, but for models with 
ages ranging from 0.25 Gyr (bottom left) to 5.25 Gyr (top left) in 
intervals of 0.5 Gyr. Error bars are shown for the model with maximum 
convective core size. These correspond to $1 \sigma$ errors in the 
frequency differences when the relative error in the individual 
frequencies is assumed to be $10^{-4}$. For this model, the radial order 
of the modes spanning the frequency range considered varies between $n=9$ 
and $n=27$. Right panel: comparison between the theoretical signal 
expected from the convective core (dashed lines) and that contained in the 
frequency differences (full curves) for models with ages ranging from 2.25 
Gyr (bottom left) to 3.75 Gyr (top left) in intervals of 0.5 Gyr. To 
determine the expected signal, we computed $\delta\nu^{\rm c}_{n,0}$ using 
Eq.~(\ref{dw2}) and then normalized it by the large separations to 
reproduce the right hand side of Eq.~(\ref{signal}).}
\label{comparison}
\end{figure*}
%%%%%%%%%%%%%%%%%%%%%%%%%%%%%%%%%%%%%%%%%%%%%%%%%%%%%%%%%%%%%%%%%%%%%%%%%
   
With this in mind, we start by testing the last assumption made when 
deriving Eq.~(\ref{signal})---namely, the weak dependence of the value of 
$D_\delta$ in the unperturbed model on the pair of modes considered. To 
this end, we compute eigenfrequencies for our youngest model. At an age of 
0.25~Gyr the convective core is extremely small, and there is no apparent 
discontinuity in the sound speed (see the top curve in 
Figure~\ref{sspeed}). Under these conditions, we expect 
$\delta\nu_{n,0}^{c}\approx 0$. Consequently, if the quantity $D_\delta$ 
were strictly independent of the pair of modes considered, the frequency 
differences defined by the expression on the left hand side (LHS) of 
Eq.~(\ref{signal}) would be approximately zero for all frequencies.

Figure~\ref{difnocc} shows the frequency differences obtained for our 
youngest model. Clearly, in the lower frequency domain the frequency 
differences deviate significantly from zero. This reflects the fact that 
$D_\delta$ is nearly independent of the pair of modes considered only in 
the high-frequency asymptotic regime. At lower frequencies this 
assumption breaks down, and consequently so does the approximation made 
when deriving Eq.~(\ref{signal}) from Eqs.~(\ref{ss02}) and (\ref{ss13}). 
Nevertheless, at high frequencies ($\ga 2$~mHz in Figure~\ref{difnocc}) 
the frequency differences converge to a constant small value, which we 
shall designate $\nu_{\rm asymp}$, confirming our assumption of the near 
degeneracy of $D_\delta$ with respect to the pairs of modes $\ell=0,2$ and 
$\ell=1,3$.

To compare these results with those obtained for older models, we 
determine the frequency differences defined by the LHS of 
Eq.~(\ref{signal}) for our sequence of models, with ages ranging from 
0.25~Gyr to 5.25~Gyr. The results are shown in the left panel of 
Figure~\ref{comparison}. It is clear from this figure that in the 
high-frequency domain, where Eq.~(\ref{signal}) is expected to be valid, 
the absolute value of the slope of the frequency differences increases 
with age. This can be understood as a direct consequence of the increase 
with age of the sound speed discontinuity at the edge of the convective 
core. According to Eq.~(\ref{dw2}), such an increase manifests itself 
through the amplitude $A_\delta$ by an increase in the absolute value of 
the frequency perturbation at fixed frequency. This increase is 
consequently reflected in the frequency differences defined by the LHS of 
Eq.~(\ref{signal}), and thus in the slope of the curves plotted in 
Figure~\ref{comparison}.

To determine whether the signature of the convective core on the frequency 
differences might be detected in future observations of solar-like 
pulsators, we have included for one of our models the expected 1$\sigma$ 
error bars for the frequency differences, calculated with the assumption 
that the relative error on the individual frequencies is 10$^{-4}$ 
\citep[e.g., 0.2~$\mu$Hz at 2~mHz as expected from CoRoT;][]{bag06}. It is 
evident from the figure that if individual frequencies are indeed 
determined with such precision, we will be able to not only detect the 
effect of the convective core on the frequencies, but also distinguish 
between different possible models based on this signature. Moreover, we 
see that our ability to detect this signature in real data will depend on 
the number of radial modes detected, as well as on their radial order.

We made additional assumptions, other than the one referred to at the 
beginning of this section, while deriving Eq.~(\ref{signal}). In 
particular, we assumed that the effect of the convective core on the 
frequencies of modes with degree $\ell \ge 1$ is negligible, when compared 
to the effect on the frequencies of radial modes. To check whether this 
assumption is valid, we compare the frequency differences calculated for a 
subset of our models with the theoretical expectations derived from 
Eq.~(\ref{dw2}). The results are shown in the right panel of 
Figure~\ref{comparison}. We see from the figure that in the high-frequency 
asymptotic regime the theoretical signal derived from the analysis of the 
wave equation near the inner turning point is indeed isolated when the 
oscillation frequencies are combined in the way suggested by 
Eq.~(\ref{signal}). We note that there is no model dependent free 
parameter in the theoretical expression for the frequency perturbations. 
The difference in the slopes of the theoretical curves at high-frequency, 
seen also in the frequency differences calculated from simulated data, is 
thus a direct consequence of the dependence of the amplitude $A_\delta$ on 
the increase in the sound speed at the edge of the convective core.

% FIGURE 5 %%%%%%%%%%%%%%%%%%%%%%%%%%%%%%%%%%%%%%%%%%%%%%%%%%%%%%%%%%%%%%
\begin{figure*}
\plottwo{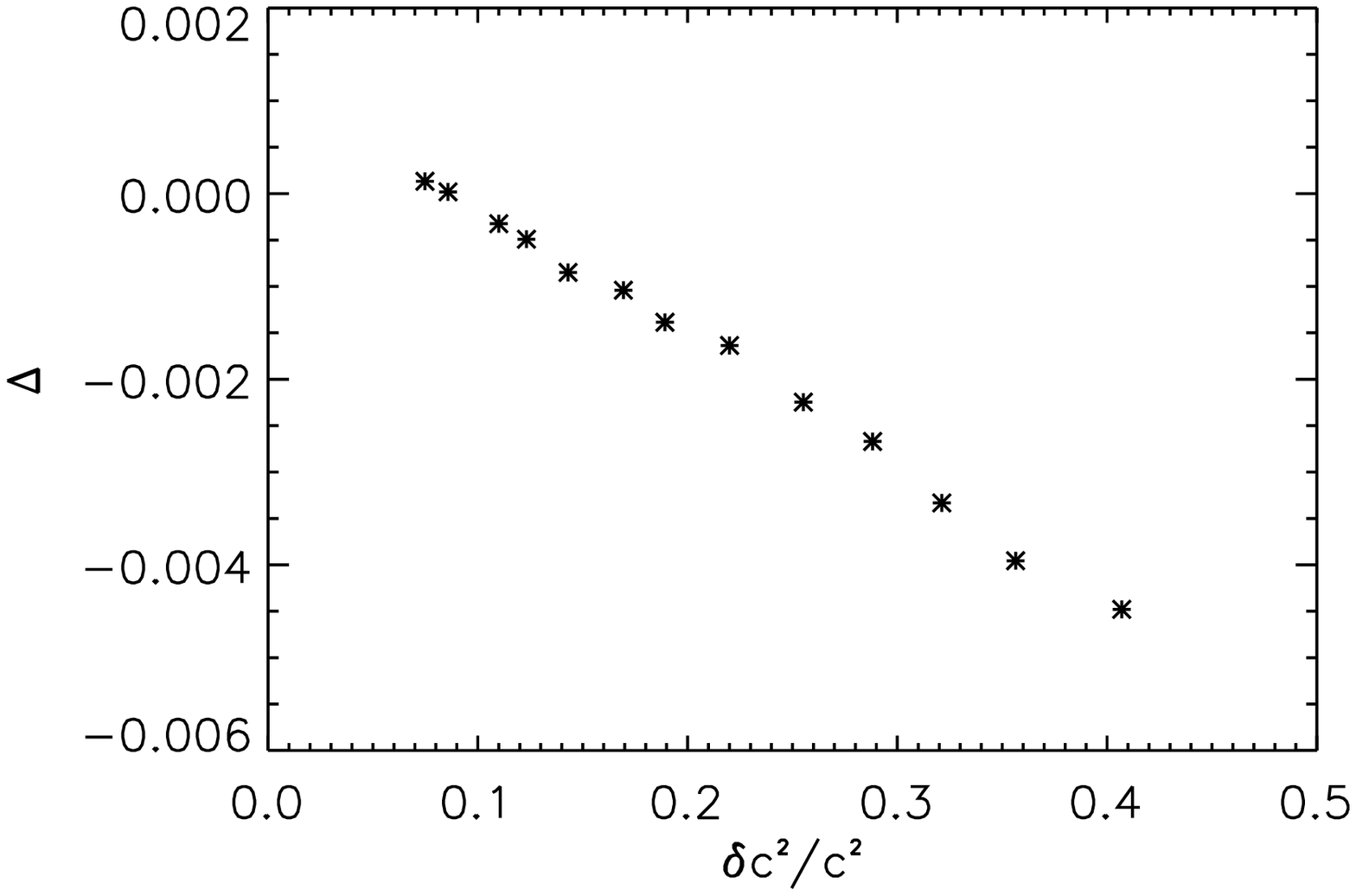}{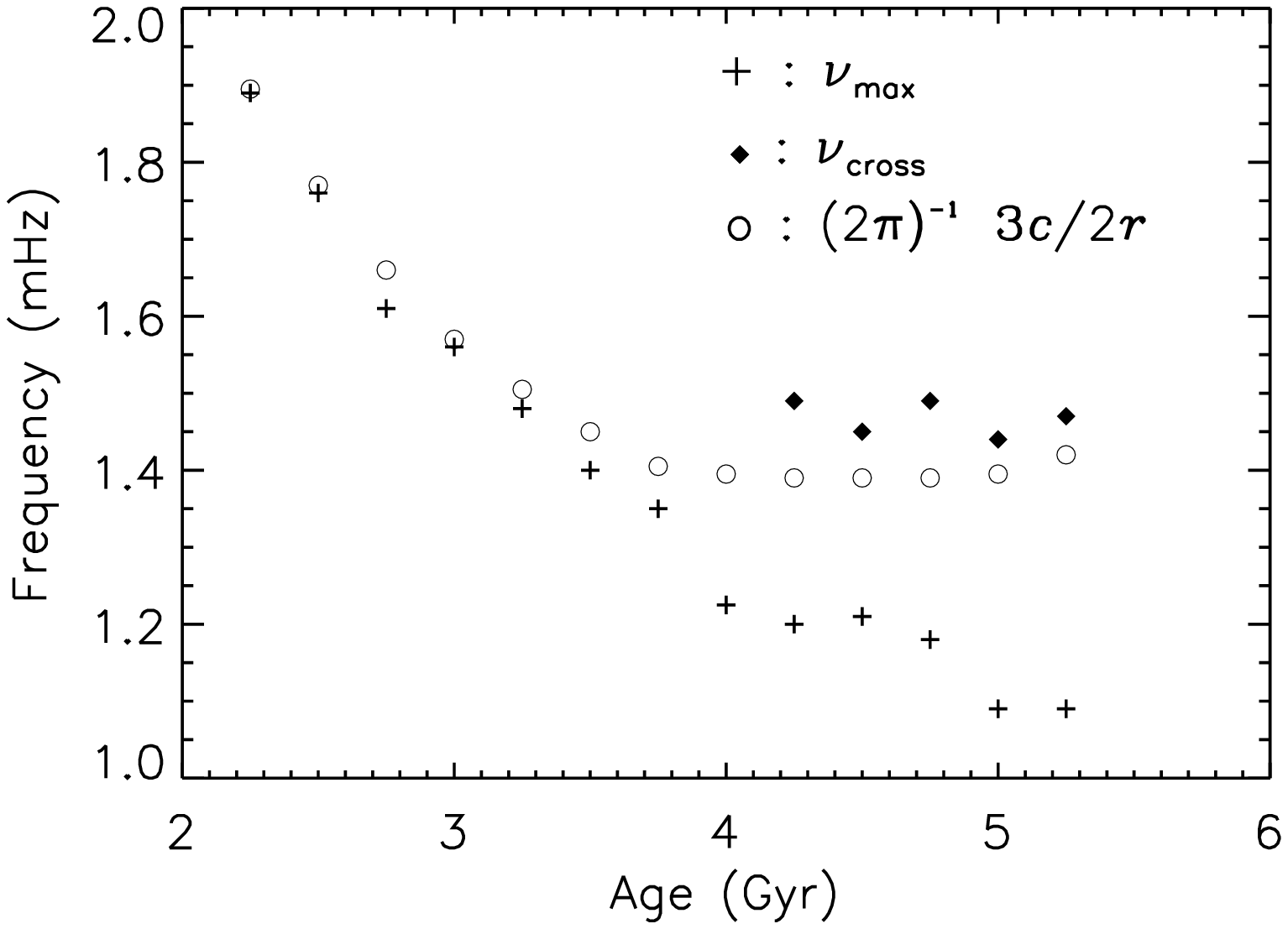}
\caption{Left panel: measurement of the slope of the frequency differences 
in the high-frequency domain as a function of the relative increase in the 
squared sound speed at the edge of the convective core (see text for the 
mathematical definition of $\Delta$). The ages of the models shown range 
from 2.25 Gyr (top left) to 5.25 Gyr (bottom right) in intervals of 
0.25~Gyr. Right panel: for models with ages ranging from 2.25 Gyr to 5.25 
Gyr, we show the frequency $\nu_{\rm max}$ at which the frequency 
differences suggested by the LHS of Eq.~(\ref{signal}) reach their maximum 
(crosses), and for the same models the value of $\,\,(2\pi)^{-1}\,3c/2r$ (circles) at the edge of the convective core. 
Since $c$ is discontinuous at the edge of the convective core, we use the 
mean of the values of $c$ on the left and right side of the discontinuity 
when calculating this expression. For models with ages greater than 4 Gyr, 
we show the frequency $\nu_{\rm cross}$ at which the frequency differences 
cross the asymptotic value for the youngest model (diamonds).}
\label{slope}
\end{figure*}
%%%%%%%%%%%%%%%%%%%%%%%%%%%%%%%%%%%%%%%%%%%%%%%%%%%%%%%%%%%%%%%%%%%%%%%%%

In contrast to these results, Figure~\ref{comparison} shows that in the 
low-frequency domain the combination of frequencies suggested by 
Eq.~(\ref{signal}) does not properly isolate the theoretical signature 
expected from Eq.~(\ref{dw2}). Just as in the case discussed for 
Figure~\ref{difnocc}, this is likely a consequence of the breakdown at low 
frequencies of the assumption that $D_\delta$ is independent of the pair 
of modes considered. According to Figure~\ref{difnocc}, when the frequency 
perturbation associated with the convective core is negligible, the 
frequency differences become negative in the low-frequency domain, 
decreasing as the frequency decreases. On the other hand, the frequency 
perturbation associated with the convective core becomes positive for 
frequencies below a particular value---namely, the frequency at which the 
position of the discontinuity in the sound speed coincides with the inner 
turning point of Eq.~(\ref{waveeq}). In the low-frequency domain, we thus 
expect our diagnostic tool to show a superposition of these two behaviors. 
Indeed, this is seen for frequencies smaller than $\sim$1.5~mHz in the 
left panel of Figure~\ref{comparison}. The competition between these two 
behaviors results in the presence of a maximum in the frequency 
differences calculated for each model.

\subsection{Characterizing the convective core}

The validation of the diagnostic tool proposed in Eq.~(\ref{signal}) in 
the high-frequency domain makes us confident that we understand the origin 
of the signature observed in the simulated data, and that we can use this 
signature to extract information about the convective core. As mentioned 
in \S\ref{theory} and \S\ref{tests}, information about the size of the 
sound speed discontinuity at the edge of the convective core can be 
extracted from the amplitude of the signal, and in principle information 
about the ratio $[c/r]_{\rm edge}$ at the edge of the convective core can 
be extracted from the value of the frequency at which the frequency 
perturbation goes through zero. Moreover, given the age dependence of the 
size of the sound speed discontinuity at the edge of the core, the 
signature discussed above might also be used to infer the evolutionary 
status of pulsating stars within the region of the H-R diagram considered 
here.

These ideas are illustrated in Figure~\ref{slope}. In the left panel the 
quantity $\Delta$, defined as the difference between the frequency 
differences $\left(\frac{D_{02}}{\Delta\nu_{n-1,1}}- 
\frac{D_{13}}{\Delta\nu_{n,0}}\right)$ of modes with radial orders $n=k+2$ 
and $n=k-2$, with $k$ such that $\nu_k \sim 1750$~$\mu$Hz for each model, is 
shown as a function of the relative increase in the squared sound speed at 
the edge of the convective core. The quantity $\Delta$ provides a measure 
of the slope of the frequency difference curves, such as those plotted in 
the left panel of Figure~\ref{comparison}, in the high-frequency domain. 
It is clear from Figure~\ref{slope} that a linear relation exists between 
$\Delta$ and the squared sound speed increase at the edge of the 
convective core. We also considered alternative quantities indicative of 
the slope of the same curves taken at constant radial order rather than at 
constant frequency. In all cases the results indicate a linear relation 
with the squared sound speed increase at the edge of the convective core. 
Note that when studying data with noise, the use of the quantity $\Delta$ 
defined by only two data points is not advisable. Instead, it would be 
more convenient to extract the slope of the frequency difference curves in 
the high-frequency domain through a least-squares fit to the appropriate 
part of the curve under consideration.

Since the proposed diagnostic tool fails to reproduce the theoretical 
signal in the low-frequency domain, the idea of using this tool to 
determine the value of the frequency at which the frequency perturbation 
associated with the convective core goes through zero---and thus 
extracting information about the physical size of the core---becomes 
compromised. Nevertheless, since the maxima that characterize the curves 
defined by the frequency differences are a consequence of the positive 
frequency perturbations in the low-frequency domain, we might expect the 
position of these maxima to contain information about the size of the 
convective core. With this in mind, we have determined for each model the 
frequency $\nu_{\rm max}$ at which the frequency differences defined by 
the LHS of Eq.~(\ref{signal}) reach their maximum value. These are plotted 
as a function of age in the right panel of Figure~\ref{slope}, and are 
compared with the frequencies $\left[(2\pi)^{-1}\,3c/2r\right]_{\rm 
edge}$ at which the frequency perturbation associated with the convective 
core goes through zero.

Clearly, the frequency at which the frequency differences reach their 
maxima, indicated by the crosses in Figure~\ref{slope}, is strongly 
correlated with the age of the model. This is not surprising since the 
magnitudes of the frequency perturbations, which influence the position of 
the maxima, are also correlated with age. Nevertheless, a change in the 
slope is noticeable around an age of $\sim$4~Gyr. This is likely a 
consequence of the contracting core. As seen in Figure~\ref{sspeed}, when 
the core begins to contract, the sound speed variation near the edge of 
the convective core begins to spread over a larger radial extent. Our 
theoretical analysis, which assumes that the sound speed variation at the 
edge of the core is well represented by a modified step function, becomes 
less adequate in this case. Nevertheless, an inspection of the frequency 
differences shown in the left panel of Figure~\ref{comparison} indicates 
that in the older models where this effect becomes evident, the amplitude 
of the frequency differences in the low-frequency domain is indeed 
substantially larger than in all other models. This may result from a 
sudden increase in the amplitude of the frequency perturbation in the 
low-frequency domain, or from an improvement in the degeneracy of 
$D_\delta$ calculated for pairs of modes with degrees $\ell=0,2$ and 
$\ell=1,3$. Whatever the case, the relative importance of the latter in 
determining the functional form of the frequency differences appears to be 
smaller for models with ages greater than $\sim$4~Gyr. Thus, one might 
expect that for these models the frequency 
$\left[(2\pi)^{-1}\,3c/2r\right]_{\rm edge}$ will be better estimated 
by the frequency $\nu_{\rm cross}$ at which the frequency differences 
cross the asymptotic value for the youngest model in Figure~\ref{difnocc}. 
Indeed this seems to be the case, as shown by the diamonds in the right 
panel of Figure~\ref{slope}.

%%%%%%%%%%%%%%%%%%%%%%%%%%%%%%%%%%%%%%%%%%%%%%%%%%%%%%%%%%%%%%%%%%%%%%%%%%%

\section{\label{DISC}Discussion}

We have derived the expected signature of small convective cores on the 
oscillation frequencies of solar-like pulsators, we have suggested a 
diagnostic tool to isolate this signature, and finally we have discussed 
what information the signature may reveal about the convective cores 
themselves and about the evolutionary status of the corresponding stars.

The results demonstrate that our ability to isolate the convective core 
signature with the suggested diagnostic tool is significantly better in 
the high-frequency asymptotic domain than at lower frequencies. 
Consequently, the information inferred from this diagnostic tool at high 
frequencies is expected to be more robust. In particular, this is the case 
for the relation found between the slope of the frequency difference 
curves and the relative increase in the squared sound speed at the edge of 
the convective core. Based on the theoretical signal, which in the 
high-frequency domain is well represented by the proposed diagnostic tool, 
we expect this relation to hold also in a more general context when 
models of different mass, chemical composition, or with different amounts 
of core overshoot are considered.

By contrast, information inferred from the behaviour of the diagnostic 
tool at lower frequencies, such as the estimation of the ratio $[c/r]_{\rm 
edge}$ from the position of the maximum in the frequency differences, or 
from the value of the frequency at which these differences cross $\nu_{\rm 
asymp}$, is likely to show a stronger model dependence. As emphasized
earlier, at lower frequencies the functional form of the diagnostic tool 
is influenced by the deviation from degeneracy of the quantity $D_\delta$. 
Consequently, this tool does not capture, as well as it does at higher 
frequencies, the effect of the convective core on the oscillations, even 
though it is still influenced by it. Therefore, it is important to extend 
the present study to different models, varying quantities such as mass, 
chemical composition, and core overshoot, to determine the extent to which 
the quantities studied---particularly in the low-frequency domain---depend 
on the properties of the models.

One aspect of major importance that has been considered by most works 
involving diagnostic tools to study the inner regions of stars is the 
ability of these tools to determine the stellar age. The correlations 
between the quantities studied in \S\ref{DATA} and the age of the models 
are clear. Hence, in principle the former might be used to infer the 
evolutionary status of pulsating stars. 
In fact, this correlation is responsible for the one found by \cite{maz06} 
between their quantity $\theta$ and the central hydrogen abundance $X_c$, 
(cf. figure 7 of their paper). Their $\theta$ is proportional to the 
difference $(D_{13}-D_{02})$ averaged over several radial orders $n$, and 
through $X_c$ it reflects the change in the slope seen in the left panel 
of our Figure~\ref{comparison}. Thus, our analysis also provides the 
theoretical basis for the correlation found by \cite{maz06}.
 Nevertheless, it is important to investigate how these correlations
 depend on other stellar parameters, and to ascertain whether there
 are degeneracies in the age determination when the appropriate
 parameter space is explored.
Also, as mentioned in \S~\ref{models}, the models used to test our 
diagnostic tool did not include diffusion. The presence of diffusion will 
likely result in a smoother transition in $c^2$. We hope to tackle the 
consequences of including diffusion in future work.

Asteroseismic data with the precision necessary to test our diagnostic 
tool observationally are soon expected from the CoRoT satellite 
\citep{bag06}. The second scheduled 5-month run will include the star 
HD~49933, which exhibits solar-like oscillations \citep{mos05} and has 
approximately the required mass and age to reveal the expected signal. 
Future observations of the slightly more massive spectroscopic binary star 
12~Bo{\"o}tis \citep{mig07} could also provide an interesting test of our 
predictions.

%%%%%%%%%%%%%%%%%%%%%%%%%%%%%%%%%%%%%%%%%%%%%%%%%%%%%%%%%%%%%%%%%%%%%%%%%%%

\acknowledgments 

The authors are very grateful to J{\o}rgen Christensen-Dalsgaard and 
Douglas Gough for helpful comments and suggestions. This work was 
supported in part by the EC's FP6, FCT and FEDER (POCI2010) through the 
HELAS international collaboration and through the project 
POCI/CTE-AST/57610/2004, by a Fulbright grant under the Mutual Educational 
Exchange Program, and by NCAR through the ECSA and HAO Visiting Scientist 
Programs. The National Center for Atmospheric Research is a federally 
funded research and development center sponsored by the U.S. National 
Science Foundation.

%%%%%%%%%%%%%%%%%%%%%%%%%%%%%%%%%%%%%%%%%%%%%%%%%%%%%%%%%%%%%%%%%%%%%%%%%%%

\appendix

\section{\label{APA}A.~Solution to the wave equation near the inner turning 
point}

In this appendix we derive an approximate solution to Eq.~(\ref{waveeq}), 
valid near the inner turning point $z_1$ at which $K^2=0$. We closely 
follow the derivation presented in \S4.8.5 of \citet{gough93}, though the 
dependent and independent variables in our Eq.~(\ref{waveeq}) differ from 
those in \citeauthor{gough93}'s Eq.~(4.8.5).

We start by defining new independent and dependent variables, $x$ and 
$\psi$ respectively
\begin{eqnarray}
x & = &\sgnk2\left[\sgnk2\frac{3}{2}
\int_{z_1}^z|K|\rmd z\right]^{\frac{2}{3}},
\label{defx} \\
\psi & = & |x|^{-\frac{1}{4}}|K|^{\frac{1}{2}}\Xi
\equiv s^{-\frac{1}{2}}\Xi.
\label{defpsi}
\end{eqnarray}
With these definitions, $x=0$ at the inner turning point. Substituting 
these into Eq.~(\ref{waveeq}) we find
\begin{equation}
\frac{\rmdd\psi}{\rmd x^2}+x\psi=-s^{\frac{3}{2}}
\frac{\rmdd s^{\frac{1}{2}}}{\rmd z^2}\psi,
\label{airy}
\end{equation}
where we keep the independent variable $z$ in the right hand side (RHS) to 
keep the equation in a condensed form.

To find an asymptotic solution to Eq.~(\ref{airy}) we regard the term on 
the RHS as a small perturbation to the Airy equation \citep{olver74}. We 
note that our original transformation of the independent variable $z={\rm 
ln}(r/r_{\rm ref})$, was motivated precisely by the need to keep the RHS 
bounded when $r$ tends to zero \citep{langer37}. In fact, a 
straightforward analysis of this term shows that it will tend to zero as 
$r$ tends to zero.

To leading order, we thus have that the solution to Eq.~(\ref{airy}) 
around the inner turning point can be obtained from
\begin{equation}
\psi\approx aA_{i}\left(-x\right)+bB_{i}\left(-x\right),
\end{equation}
where $A_i$ and $B_i$ are the solutions to the Airy equation, and $a$ and 
$b$ are constants. Since the solution must tend to zero as $x$ tends to 
$-\infty$, we must have $b=0$. Using the ascending series for $A_i$ 
\citep[e.g.][\S10.4, p.446]{abramowitz72}, we thus find for $|x|\ll 1$
\begin{equation}
\psi\approx a\left[A_i\left(0\right)-A_i^\prime\left(0\right)x 
+\mathcal{O}\left(x^3\right)\right],
\end{equation}
where 
\begin{eqnarray}
A_i\left(0\right)=\frac{1}{3^{\frac{2}{3}}\Gamma\left(\frac{2}{3}\right)} 
&\,\,\, {\rm and}\,\,\,& 
A_i^\prime\left(0\right)=
-\frac{1}{3^{\frac{1}{3}}\Gamma\left(\frac{1}{3}\right)}
\nonumber,
\end{eqnarray}
and $\Gamma$ is the Gamma function. 
Transforming back, we then find that
\begin{eqnarray}
\Xi \approx  a\, s^{\frac{1}{2}}\left[A_{\rm i}\left(0\right)- 
{A^\prime}_{\rm i}\left(0\right)x\right],
\end{eqnarray}
with $s$ defined by Eq.~(\ref{defpsi}).

%%%%%%%%%%%%%%%%%%%%%%%%%%%%%%%%%%%%%%%%%%%%%%%%%%%%%%%%%%%%%%%%%%%%%%%%%%%

\section{\label{APB}B.~Variational analysis}

In this appendix we derive the expression for the frequency perturbations 
resulting from the presence of a small convective core. Our starting point 
is Eq.~(\ref{vpp}), with $\delta I_1$ and $\delta I_2$ defined by 
Eqs.~(\ref{di1}) and (\ref{di2}). Introducing the new variables $x$ and 
$\psi$ into Eqs.~(\ref{di1}) and (\ref{di2}) and noting that
\begin{eqnarray}
\left(\frac{\rmd\psi}{\rmd x}\right)^2=
\frac{1}{2}\frac{\rmdd\psi^2}{\rmd x^2}-\psi\frac{\rmdd\psi}{\rmd x^2} 
& \,\,\, {\rm and} \,\,\, & 
\frac{\rmdd\psi}{\rmd x^2}\approx -x\psi,
\end{eqnarray}
we find, after some algebra,
\begin{eqnarray}
2I_1\omega\delta\omega \approx 
\int_{X_0}^{X_R}\left[f_1\psi^2+f_2\frac{\rmd\psi^2}{\rmd x}
+f_3\frac{\rmdd \psi^2}{\rmd x^2}\right]\rmd x,
\label{dw1}
\end{eqnarray}
where $X_0$ and $X_R$ are the limits of the original integral in terms of 
the new variable $x$ and
\begin{eqnarray}
f_1&=&r^3q^2\left\{s^{-1}\left[x+\left(\frac{1}{q} \frac{\rmd q}{\rmd 
x}\right)^2\right]\delta\left(\gamma p\right)- \frac{\rmd}{\rmd 
x}\left[3\delta\left(\gamma p\right)-4\delta p\right]- 
\omega^2r^2s\delta\rho\right\},
\nonumber\\
f_2&=&\frac{1}{2}r^3s^{-1}\frac{\rmd q^2}{\rmd x}\delta\left(\gamma 
p\right),
\nonumber \\
f_3&=&\frac{1}{2}r^3s^{-1}q^2\delta\left(\gamma p\right),
\nonumber 
\end{eqnarray}
with $q=s^{1/2}r^{-3/2}{r_{\rm ref}}^{-1/2}\rho^{-1/2}c^{-1}$.

Next, we express Eq.~(\ref{dw1}) in terms of the perturbations 
$\delta(\gamma p)/\gamma p$ and $\delta c^2/c^2$. Using the relations
\begin{eqnarray}
\frac{\delta\rho}{\rho}=\frac{\delta\left(\gamma p\right)}{\gamma 
p}-\frac{\delta c^2}{c^2} & \,\,\, {\rm and}\,\,\, &\frac{\rmd\delta 
p}{\rmd x}=-\rho grs\left(\frac{\delta g}{g}+\frac{\delta\left(\gamma 
p\right)}{\gamma p}-\frac{\delta c^2}{c^2}\right),
\end{eqnarray}
we find
\begin{eqnarray}
2I_1\omega\delta\omega \approx 
\int_{X_0}^{X_R}\left[\left(\delta\mF_1+\frac{\rmd}{\rmd 
x}\left(\delta\mF_0\right)\right)\psi^2+\delta\mF_2\frac{\rmd\psi^2}{\rmd 
x}+\delta\mF_3\frac{\rmdd \psi^2}{\rmd x^2}\right]\rmd x,
\label{dw3}
\end{eqnarray}
where
\begin{eqnarray}
\delta\mF_0&=&-\left(3r^3q^2\gamma p\right)\frac{\delta\left(\gamma 
p\right)}{\gamma p}, 
\nonumber \\
\delta\mF_1&=&\left\{r^3q^2\left[s^{-1}\left(x+\left(\frac{1}{q}\frac{\rmd 
q}{\rmd x}\right)^2\right)\gamma p-\rho 
rs\left(4g+\omega^2r\right)\right]+3\gamma 
p\frac{\rmd\left(r^3q^2\right)}{\rmd x}\right\}\frac{\delta\left(\gamma 
p\right)}{\gamma p} \nonumber \\ &&+\left[r^4q^2\rho 
s\left(4g+\omega^2r\right)\right]\frac{\delta c^2}{c^2} \nonumber \\ & 
&-\left(4r^4q^2\rho sg\right)\frac{\delta g}{g}, 
\nonumber \\
\delta\mF_2&=&\left(\frac{1}{2}r^3s^{-1}\frac{\rmd q^2}{\rmd x}\gamma 
p\right)\frac{\delta\left(\gamma p\right)}{\gamma p}, 
\nonumber \\
\delta\mF_3&=&\left(\frac{1}{2}r^3s^{-1}q^2\gamma 
p\right)\frac{\delta\left(\gamma p\right)}{\gamma p}.
\nonumber
\end{eqnarray}
 
The perturbation associated with the discontinuity at the edge of the 
convective core is rather localized. Thus, we consider an interval 
$[X_a,X_b]$ in which the perturbation differs from zero and assume it to 
be equal to zero outside that interval. 
By requiring both that the sound speed is everywhere the same in the 
perturbed and unperturbed models except near the discontinuity, and that 
both models are in hydrostatic equilibrium, the small difference in their 
density profiles extends beyond the region of the discontinuity, towards 
the center of the star. Nevertheless, because we are interested in 
isolating the frequency perturbation associated with the rapid variation 
at the edge of the core, we do not consider the perturbation arising from 
this additional difference between the two models.

Integrating by parts all terms of 
the integral on the RHS of Eq.~(\ref{dw3}) except for that involving 
$\delta\mF_0$, we find
\begin{eqnarray}
2I_1\omega\delta\omega \approx 
-\int_{X_a}^{X_b}\left[\left(\frac{\rmd\delta\mF_1}{\rmd 
x}\right)\overline{\psi^2}+\left(-\frac{\rmd\delta\mF_0}{\rmd 
x}+\frac{\rmd\delta\mF_2}{\rmd x}\right)\psi^2+\frac{\rmd\delta\mF_3}{\rmd 
x}\frac{\rmd\psi^2}{\rmd x}\right]\rmd x,
\label{dw4}
\end{eqnarray}
where $\overline{\psi^2}=\int\psi^2\rmd x$. 

At the edge of the convective core, $\delta c^2 /c^2$ is discontinuous 
while all other perturbations in Eq.~(\ref{dw3}) are not. Hence, the main 
contribution to the frequency perturbations will come from the term 
involving the perturbation to the sound speed. Accordingly, we neglect all 
other terms and find
\begin{eqnarray}
2I_1\omega\delta\omega \approx -\int_{X_a}^{X_b}\frac{\rmd}{\rmd 
x}\left[r^4q^2\rho s\left(4g+\omega^2r\right)\frac{\delta 
c^2}{c^2}\right]\overline{\psi^2}\rmd x.
\label{dw5}
\end{eqnarray}

We model the sharp variation $\delta c^2 /c^2$ at the edge of the 
convective core, $x_{\rm d}$, with a function of the type $\delta c^2 /c^2 
\propto [H(x-x_{\rmd})-1/2]$, where $H$ is the Heaviside Step Function and 
the proportionality constant is determined by the amplitude of the sharp 
variation. Under this assumption, and keeping in mind that the term 
multiplying $\delta c^2 /c^2$ in Eq.~(\ref{dw5}) is a slowly varying 
function of $x$, we have
\begin{equation}
\frac{\rmd}{\rmd x}\left[r^4q^2\rho s\left(4g+\omega^2r\right)\frac{\delta 
c^2}{c^2}\right]\approx A_\delta \left[r^4q^2\rho 
s\left(4g+\omega^2r\right)\right]\delta\left(x-x_\rmd\right),
\label{deriv}
\end{equation}
where $A_\delta$ is the amplitude associated with the $\delta$-function. 
We thus find, after some algebra,
\begin{eqnarray}
2I_1\omega\delta\omega \approx -A_\delta 
a^2A_i\left(0\right)^2\left[\frac{r\left(4g+\omega^2r\right)} {r_{\rm 
ref}c^2|K|^2}\left(|x|x-\frac{A^\prime_{\rm i}(0)}{A_{\rm 
i}(0)}|x|x^2+\frac{1}{3} \frac{A^\prime_{\rm i}(0)^2}{A_{\rm 
i}(0)^2}|x|x^3\right)\right]_{x=x_{\rm d}}.
\label{dw6}
\end{eqnarray}
In practice, if the approximations introduced above are reasonably 
satisfied, $A_\delta$ is expected to be proportional to the amplitude of 
the sharp variation $\delta c^2 /c^2$ at $x_\rmd$, with the 
proportionality constant independent of the model considered. Indeed we 
found this to be the case in all of our models except for those older than 
$\sim$4~Gyr. For these models the discontinuity in $c^2$ is not as sharp 
as for younger models. Even so, the functional form of the perturbations 
was still found to be relatively well represented by Eq.~(\ref{dw6}) in 
this case.

Finally, the constant $a$ can be expressed in terms of the amplitude of 
the eigenfunction $\xi$ at the inner turning point. Combining the 
definitions of $\psi$ and $\Xi$, we find
\begin{eqnarray}
\left[\psi\right]_{x=0}=\left[q^{-1}\xi\right]_{x=0}=a A_i\left(0\right).
\end{eqnarray} 
Substituting this expression for $a$ into Eq.~(\ref{dw6}), we finally 
have
\begin{eqnarray}
\delta\omega \approx -\frac{A_\delta}{2\omega I_1}\left[\frac{|K|r^3\rho 
c^2\xi^2}{|x|^{1/2}}\right]_{x=0}\left[\frac{r\left(4g+\omega^2r\right)} 
{c^2|K|^2}\right]_{x=x_{\rm d}}\left[|x|x-\frac{A^\prime_{\rm 
i}(0)}{A_{\rm i}(0)}|x|x^2+\frac{1}{3} \frac{A^\prime_{\rm i}(0)^2}{A_{\rm 
i}(0)^2}|x|x^3\right]_{x=x_{\rm d}}.
\label{final}
\end{eqnarray}
For practical reasons, when calculating the theoretical frequency 
perturbations from Eq.~(\ref{final}), we use the quantities $\omega$, 
$I_1$ and $\xi_{x=0}$ calculated for the model with the discontinuity. To 
first order, this should have no impact on the theoretical signal.

%%%%%%%%%%%%%%%%%%%%%%%%%%%%%%%%%%%%%%%%%%%%%%%%%%%%%%%%%%%%%%%%%%%%%%%%%%%

\end{document}